\documentclass{ws-ijmpb}

\begin{document}

\markboth{Michel Planat} {Huyghens, Bohr, Riemann and Galois:
Phase-Locking}

%
\catchline{}{}{}{}{}
%

\title{ HUYGHENS, BOHR, RIEMANN AND GALOIS: PHASE-LOCKING}

\author{MICHEL PLANAT}

\address{Institut FEMTO-ST, Departement LPMO \\32 Avenue de
l'Observatoire, 25044 Besan\c{c}on Cedex, France\\
planat@lpmo.edu}



\maketitle

\begin{history}
\received{Day Month Year}
\revised{Day Month Year}
\end{history}

\begin{abstract}
Several mathematical views of phase-locking are developed. The
classical Huyghens approach is generalized to include all harmonic
and subharmonic resonances and is found to be connected to $1/f$
noise and prime number theory. Two types of quantum phase-locking
operators are defined, one acting on the rational numbers, the
other on the elements of a Galois field. In both cases we analyse
in detail the phase properties and find them related respectively
to the Riemann zeta function and to incomplete Gauss sums.
\end{abstract}


\section{Introduction}

In the crude sense phase--locking occurs whenever the erratic
behavior of one single piece shifts to the ordered behavior of the
whole system. There is a huge number of phase-locked systems
(populations of crickets, yeast cells, lasers ...) and none
universal mechanism should be expected. The concept of phase
pervades the whole physics and we will show that its mathematical
counterpart touches several intriguing open problems.

Working experimentally it was found that the interleaving of
frequencies and phases of electronic oscillators interacting in
non linear circuits follows arithmetical rules. Continued fraction
expansions, prime number decompositions and related number
theoretical concepts were successfully used to account for the
experimental effects in mixers and phase-locked
loops.\cite{FNL01,APL02} We also made use of these tools within
the field of quantum optics emphasizing the hidden connection
between phase-locking and cyclotomy.\cite{PLA03} Finally a class
of optimal states in quantum information happens to be quantum
phase states constructed (phase-locked) from Galois fields and
rings.\cite{Planat05} In this paper we show that their properties
are related to incomplete Gauss sums.




\section{Classical Phase-Locking: from Huyghens to the Prime Numbers}
\label{Classical}

{\it Being obliged to stay in my room for several days and also
occupied in making observations on my two newly made clocks, I have
noticed a remarkable effect which no one could have ever thought of.
It is that these two clocks hanging next to one another separated by
one or two feet keep an agreement so exact that the pendulums
invariably oscillate together without variation. After admiring this
for a while, I finally figured out that it occurs through a kind of
sympathy: mixing up the swings of the pendulums, I have found that
within a half hour always return to consonance and remain so
constantly afterwards as long as I let them go. I then separated
them, hanging one at the end of the room and the other fifteen feet
away, and noticed that in a day there was five seconds difference
between them. Consequently, their earlier agreement must in my
opinion have been caused by an imperceptible agitation of the air
produced by the motion of the pendulums.}

 The citation is taken from Ref.~\refcite{LaserPhysics74}. The authors cite a
later letter by Huyghens that the coupling mechanism was in fact a
small vibration transmitted through the wall, and not movement of
air:

 {\it Lord Rayleigh (1907) made similar observations about two
driven tuning forks coupled by vibrations transmitted through the
table on which both forks sat... Locking in triode circuits was
explained by Van der Pol (1927) who included in the equation for the
triode oscillator an external electromotive force as given in
\begin{equation}
\frac{d^2v}{dt^2}-\frac{d}{dt}(gv-\beta' v^3)+\omega^2 v=\omega_0^2
V_0 \sin \omega_0 t,
\end{equation}
where $g$ is the linear net gain (i.e. the gain in excess of losses,
$\beta'$ the saturation coefficient, and $\omega$ is the resonance
frequency in the absence of dissipation or gain. He showed that when
an external electromotive force is included, of frequency
$\omega_0$, and tuned close to the oscillator frequency $\omega$,
the oscillator suddenly jumped to the external frequency. It is
important to note that the beat note between the two frequencies
vanishes not because the two frequencies vanish, not because the
triode stops oscillating, but because it oscillates at the external
frequency.

We can show the locking effect by utilizing the slowly varying
amplitude approach, including a slowly varying phase $\Phi$ and
oscillation at the external frequency $\omega_0$ and amplitude $V$
\begin{equation}
\frac{d\Phi}{dt}+K \sin \Phi= \omega-\omega_0=\omega_{LF},
\label{Adler}
\end{equation}
where we use $\omega_{LF}$ for the detuning term and $K=\omega_0
V_0/V$ for the locking coefficient}.\cite{LaserPhysics74}

The regime just described is the so-called injection locking
regime, also found in injection-locked lasers. The equation
(\ref{Adler}) is the so-called Adler's equation of
electronics.\cite{Adler46}

One way to synthesize (\ref{Adler}) is through to the phase-locked
loop of a communication receiver. The receiver is designed to
compare the information carrying external oscillator (RF) to a
local oscillator (LO) of about the same high frequency through a
non linear mixing element. For narrow band demodulation one uses a
discriminator of which the role is first to differentiate the
signal, that is to convert frequency modulation (FM) to amplitude
modulation (AM) and second to detect its low frequency envelope:
this is called baseband filtering. For more general FM
demodulation one uses a low pass filter instead of the
discriminator to remove the high frequency signals generated after
the mixer. In the closed loop operation a voltage controlled LO
(or VCO) is used to track the frequency of the RF. Phase
modulation is frequently used for digital signals because low bit
error rates can be obtained despite poor signal to noise ratio in
comparison to frequency modulation.\cite{Klapper72}

Let us consider a type of receiver which consists in a mixer, in the
form of a balanced Schottky diode bridge and a low pass filter. If
$f_0$ and $f$ are the frequencies of the RF and the LO, and
$\theta(t)$ and $\psi(t)$ their respective phases, the set mixer and
filter essentially behaves as a phase detector of sensitivity $u_0$
(in Volts/rad.), that is the instantaneous voltage at the output is
the sine of the phase difference at the inputs
\begin{equation}
u(t)=u_0\sin(\theta(t)-\psi(t)).
\end{equation}
The nonlinear dynamics of the set-up in the closed loop
configuration is well described by introducing the phase
difference $\Phi(t)=\theta(t)-\psi(t)$. Using
$\dot{\theta}=\omega_0$ and $\dot{\psi}(t)=\omega+Au(t)$, with
$\omega_0=2\pi f_0$, $\omega=2\pi f$ and A (in rad. Hz/Volt) as
the sensitivity of the $VCO$, one recovers Adler's equation
(\ref{Adler}) with the open loop gain $K=u_0A$. Such a set up is
called a phase-locked loop (or PLL).

Equation (\ref{Adler}) is integrable but its solution looks
complex.\cite{DosSantos98} If the frequency shift $\omega_{LF}$
does not exceed the open loop gain $K$, the average frequency
$\langle \dot{\Phi} \rangle$ vanishes after a finite time and
reaches the stable steady state
$\Phi(\infty)=2l\pi+\sin^{-1}(\omega_{LF}/K)$, $l$ integer. In
this phase-tracking range of width $2K$ the $RF$ and the $LO$
oscillators are also frequency-locked. Outside the mode-locking
zone there is a sech shape beat signal of frequency
\begin{equation}
\tilde{\omega}_{LF}=\langle \dot{\Phi}(t)
\rangle=(\omega_{LF}^2-K^2)^{1/2}. \label{lowfrequency}
\end{equation}
\begin{figure}[htbp]
\centering{\resizebox{7cm}{!} {\includegraphics{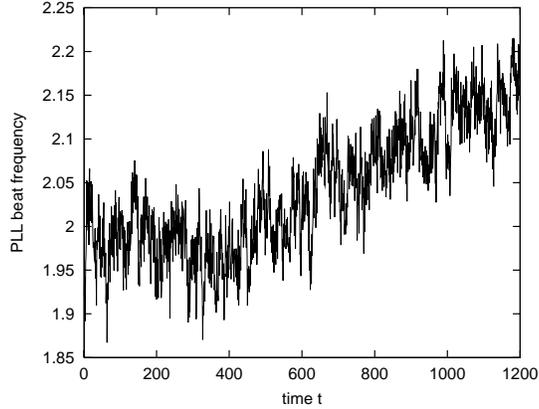}}}
\caption{Fluctuating counts of the beat frequency (in $Hz$) close to
the phase-locked zone. The inputs are quartz oscillators at $10$
MHz. The power spectrum has a pure $1/f$ dependance}
\end{figure}
The sech shape signal and the nonlinear dependance on parameters
$\omega_{LF}$ and $K$ are actually found in
experiments.\cite{DosSantos98,APL02} In addition the frequency
$\omega_{LF}$ is fluctuating (see Fig. 1). It can be characterized
by the Allan variance $\sigma^2(\tau)$ which is the mean squared
value of the relative frequency deviation between adjacent samples
in the time series, averaged over an integration time $\tau$.
Close to the phase-locked zone the Allan deviation is
\begin{equation}
\sigma(\tau)=\frac{\sigma_0 K}{\tilde{\omega}_{LF}},
\label{1surfnoise}
\end{equation}
where $\sigma_0$ is a residual frequency deviation depending of
the quality of input oscillators and that of the phase detector.
Allan deviation is found independent of $\tau$ which is a
signature of a $1/f$ frequency noise of power spectral density
$S(f)=\sigma/(2 \ln 2 f)$. One way to predict the dependence
(\ref{1surfnoise}) is to use differentiation of
(\ref{lowfrequency}) with respect to the frequency shift
$\tilde{\omega}_{LF}$ so that
\begin{equation}
\delta\tilde{\omega}_{LF}=\delta\omega_{LF}
(1+K^2/\tilde{\omega}_{LF}^2)^{1/2}.\label{deltaLF}
\end{equation}
Relation (\ref{deltaLF}) is defined outside the mode-locked zone
$|\omega_{LF}|>K$; close to it, if the effective beat note
$\tilde{\omega}_{LF}\le K$, the square root term is about
$K/\tilde{\omega}_{LF}$. If one identifies $\delta
\omega_{LF}/\tilde{\omega}_{LF}$ with a bare Allan deviation
$\sigma_0$ and $\delta \tilde{\omega}_{LF}/\tilde{\omega}_{LF}$
with a magnified Allan deviation $\sigma$ one explains the
experimental result (\ref{1surfnoise}). One can conclude that,
either the PLL set-up behaves as a microscope of an underlying
flicker floor $\sigma_0$, or the $1/f$ noise is some dynamical
property of the PLL. In the past we looked at a possible low
dimensional structure of the time series and found a stable
embedding dimension lower or equal to $4$.\cite{IEEE00} But at
that time the dynamical model of $1/f$ noise still remained
elusive.

Adler's model presupposes a fundamental interaction
$\omega_{LF}=|\omega_0-\omega(t)|$ in the mixing of the two input
oscillators. But the practical operation of the phase detector
involves harmonic interactions of the form $\omega_{LF}=|p
\omega_0 -q \omega(t)| \le \omega_c=2\pi f_c$, where $p$ and $q$
are integers and $f_c$ is the cut-off frequency of the low pass
filter. This can be rewritten by introducing the frequency ratios
$\nu=\frac{\omega (t)}{\omega_0}$ and
$\mu=\frac{\omega_{LF}}{\omega_0}$ as $\mu=q|\nu-\frac{p}{q}|$.
This form suggests that the aim of the receiver is to select such
pairs $(p,q)$ which realize a \lq\lq good" approximation of the
\lq\lq real" number $\nu$. There is a mathematical concept which
precisely does that: the diophantine approximator. It selects such
pairs $p_i$ and $q_i$, coprime to each other, i.e. with greatest
common divisor $(p_i,q_i)=1$ from the continued fraction expansion
of $\nu$ truncated at the index $i$
\begin{equation}
\nu=\{a_0;a_1,a_2,\cdots
a_i\}=a_0+1/(a_1+1/(a_2+1/\cdots+1/a_i))=\frac{p_i}{q_i}.
\label{confrac}
\end{equation}
The diophantine approximation satisfies
\begin{equation}
|\nu-\frac{p_i}{q_i}|\le\frac{1}{a_{i+1}q_i^2}. \label{dioph}
\end{equation}
The fraction $\frac{p_i}{q_i}$ is a so-called convergent and the
$a_i'$'s are called partial quotients. The approximation is
truncated at the index $i$ just before the partial quotient
$a_{i+1}$. It should be observed that diophantine approximations
are different from decimal approximations $\frac{c_i}{d_i}$ for
which one gets $|\nu-\frac{c_i}{d_i}|\le \frac{1}{d_i}$. It was
shown in Ref.~\refcite{FNL01}, using the filtering condition, that
$a_{i+1}$ is given by
\begin{equation}
a_{i+1}=\left[ \frac{f_0}{f_c q_i}\right], \label{filter}
\end{equation}
where $[~ ]$ denotes the integer part. For example if one chooses
$f_0=10$ MHz and $f_c=300$ kHz, the fundamental basin
$\frac{p_i}{q_i}=\frac{1}{1}$ will be truncated if $a_{i+1}\ge 33$
and the basin $\frac{p_i}{q_i}=\frac{3}{5}$ will be truncated if
$a_{i+1}\ge 6$. The resulting full spectrum is a superposition of
V-shape basins of which the edges are located at
\begin{eqnarray}
&\nu_1=\{a_0;,a_1,a_2,\cdots,a_i,a_{i+1}\},\nonumber\\
&\nu_2=\{a_0;a_1,a_2,\cdots,a_{i-1},1,a_{i+1}\}, \label{edge}
\end{eqnarray}
where the partial expansion before $a_{i+1}$ corresponds to the two
possible continued fractions of the rational number
$\frac{p_i}{q_i}$. The basin of number $\nu=\frac{3}{5}=\{0;1,1,2\}$
extends to $\nu_1=\{0;1,1,2,33\}=\frac{19}{32}\simeq 0.594 $,
$\nu_2=\{0;1,1,1,1,33\}=\frac{31}{34}\simeq 0.618$. For a reference
oscillator with $f_0=10$ MHz this corresponds to a frequency
bandwidth $(0.618-0.594).10^7$ MHZ=240 kHz.

With these arithmetical rules in mind one can now tackle the
difficult task of accounting for phase-locking of the whole set of
harmonics in the beat frequency.\cite{APL02}

%
\begin{equation}
\omega_{LF}=|p_i\omega_0-q_i\omega(t)|, \label{beat}
\end{equation}
%
%
%
Some essential features can be found in the standard Arnold map
model\cite{APL02}
\begin{equation}
\Phi_{n+1}=\Phi_n+2\pi\Omega-c~ \sin \Phi_n, \label{Arnold}
\end{equation}
where $\Omega=\frac{\omega}{\omega_0}$ is the bare frequency ratio
and $c=\frac{K}{\omega_0}$. Such a nonlinear map is studied by
introducing the winding number $\nu=\lim _{n \rightarrow \infty }
(\Phi_n-\Phi_0)/(2 \pi n)$. The limit exists everywhere as long as
$c<1$, the curve $\nu$ versus $\omega$ is a devil's staircase with
steps attached to rational values $\Omega=\frac{p_i}{q_i}$ and
width increasing with the coupling coefficient $c$. The
phase-locking zones may overlap if $c>1$ leading to chaos from
quasi-periodicity.\cite{Cvit92}

The Arnold map is also a relevant model of a short Josephson
junction shunted by a strong resistance $R$ and driven by a periodic
current of frequency $\omega_0$ and amplitude $I_0$. Steps are found
at the driving voltages $V_r=RI_0=r(\hbar\omega_0/2e)$, $r$ a
rational number. Fundamental resonances $r=n$, $n$ integer, have
been used to achieve a voltage standard of relative uncertainty
$10^{-7}$.

To appreciate the impact of harmonics on the coupling coefficient
one may observe that each harmonic of denominator $q_i$ creates
the same noise contribution $\delta \omega_{LF}=q_i \delta
\omega(t)$. They are $\phi(q_i)$ of them, where $\phi(q_i)$ is the
Euler totient function, that is the number of integers less or
equal to $q_i$ and prime to it; the average coupling coefficient
is thus expected to be $1/\phi(q_i)$. In Ref.~\refcite{APL02} a
more refined model is developed based on the properties of prime
numbers. It is based on defining a coupling coefficient as $c^*=c
\Lambda(n;q_i,p_i) $ with $\Lambda(n;q_i,p_i)$ a generalized
Mangoldt function.\cite{APL02} It is defined as
%
%
%
\begin{equation}
\Lambda(n;q_i,p_i)= \left\{\begin{array}{ll}
\ln b &~~\mbox{if}~ n=b^k,~b~\mbox{a~prime}~\rm{and}~n= p_i~ (\rm{mod}~q_i),\\
 0 &~~ \mbox{otherwise}.\\
\end{array}\right.
\label{equa4}
\end{equation}
The classical Mangoldt function is $\Lambda(n)=\Lambda(n;1,1)$. It
is thus the coupling coefficient of the fundamental resonance
$1/1$. The important result of that analysis is to exhibit a
fluctuating average coefficient as follows
\begin{equation}
c_{\rm{av}}^*/c=\frac{1}{t} \sum_{n=1}^{t}
\Lambda(n;q_i,p_i)=\frac{1}{\phi(q_i)}+\epsilon(t) ,\label{flucgain}
\end{equation}
with $\epsilon(t)=O(t^{-1/2}\ln^2(t))$ which is known to be a good
estimate as long as $q_i<\sqrt t$.\cite{FNL01} The average
coupling coefficient shows the expected dependance on $q_i$. In
addition there is an arithmetical noise $\epsilon(t)$ with a low
frequency dependance of the power spectrum reminding $1/f$ noise.
Although that stage of the theory is not the last word of the
story, it is quite satisfactory that this approach, based on
phase-locking of the full set of harmonics, is accounting for the
main aspects of $1/f$ noise found in experiments.

\section{Quantum Phase-Locking}
\label{Quantum}

{\it Apparently\footnote{One referee nominated
Madelung\cite{Madelung} as the first investigator
  of quantum phase operators.} Dirac was the first to attempt a definition of a
phase operator by means of an operator amplitude and phase
decomposition. As we have discussed, with a complex $c$-number
$a=Re^{i\Phi}$ one obtains the phase via $e^{i\Phi}=a/R$. Similarly,
he sought to decompose the annihilation operator $a$ into amplitude
and phase components... After a brief calculation we obtain a
relation indicating that the number operator $N$ and phase operator
$\Phi$ are canonically conjugate
\begin{equation}
[N,\Phi]=1.
\end{equation}
The equation immediately leads to a number-phase uncertainty
relation which is often seen
\begin{equation}
\delta N ~\delta \Phi \ge 1/2.
\end{equation}
However, all of the previous development founders upon closer
examination.}

 This is taken from Ref.~\refcite{Lynch95}, a comprehensive review of the quantum
 phase problem. See also Ref.~\refcite{Schleich93}.

 To approach the phase-locking problem within quantum mechanics one
 can start from the theory of the harmonic oscillator. The natural
 objects are the Fock states (the photon occupation states)
 $|n\rangle$ who live in an infinite dimensional Hilbert space.
 They are orthogonal to each other: $\langle n|m\rangle
 =\delta_{mn}$, where $\delta_{mn}$ is the Dirac symbol. The states form a
 complete set: $\sum_{n=0}^{\infty}|n\rangle \langle n|=1$.

 The annihilation operator removes one photon from the electromagnetic field
\begin{equation}
a|n\rangle=\sqrt{n}|n-1\rangle, n=1,2,\cdots
\end{equation}
Similarly the creation operator $a^{\dag}$ adds one photon:
$a^{\dag}|n\rangle=\sqrt{n+1}|n+1\rangle$, $n=0,1,\cdots$ There is
the commutation relation $[a,a^{\dag}]=1$. The operator $N=a
a^{\dag}$ has the meaning of the particle number operator and
satisfies the eigenvalue equation $N|n\rangle=n|n\rangle$.

Eigenvectors of the annihilation operator are the so-called
coherent states $|\alpha\rangle$\cite{Dodonov03} and are the ones
generated by  single mode laser operated well above threshold.
%
%

States of well defined phase escaping the inconsistencies of
Dirac's formulation were build by Susskind and
Glogower.\cite{Susskind64} They correspond to the eigenvalues of
the exponential operator
\begin{equation}
E=e^{i\Phi}=(N+1)^{-1/2}a=\sum_{n=0}^{\infty}|n\rangle\langle n+1|.
\end{equation}
Using the Hermitian conjugate operator $E^{\dag}=e^{-i\Phi}$, one
gets $E E^{\dag}=1$, $E^{\dag}E=1-|0\rangle\langle0|$, i.e. the
unitarity of $E$ is spoiled by the vacuum-state projector
$|0\rangle\langle0|$. The Susskind-Glogower phase states satisfy the
eigenvalue equation $E|\Psi\rangle=e^{i\psi}|\Psi\rangle$; they are
given as
\begin{equation}
|\Psi\rangle=\sum_{n=0}^{\infty}e^{in\psi}|n\rangle.
\label{Susskind}
\end{equation}
Like the coherent states the phase states are non orthogonal and
they form an overcomplete basis which solves the identity
operator: $\frac{1}{2\pi}\int_{-\pi}^{\pi}d\psi|\psi\rangle\langle
\psi|=1$. The operator $\cos \Phi=\frac{1}{2}(E+E^{\dag})$ is used
in the theory of Cooper pair box with a very thin junction when
the junction energy $E_J \cos \Phi$ is higher than the
electrostatic energy.\cite{Bouchiat98}

Further progress in the definition of phase operator was obtained
by Pegg and Barnett.\cite{Pegg89} The phase states are defined
from the discrete Fourier transform (or more precisely the quantum
Fourier transform since the superposition is on Fock states not on
real numbers)
\begin{equation}
|\theta_k\rangle=\frac{1}{\sqrt{q}}\sum_{n=0}^{q-1}\exp(2i\pi\frac{k}{q}n)|n\rangle.
\label{QFT}
\end{equation}
The states are eigenstates of the Hermitian phase operator
\begin{equation}
\Theta_q=\sum_{k=0}^{q-1}\theta_k |\theta_k\rangle\langle
\theta_k|,\label{Pegg} \end{equation}
with $\theta_k =\theta_0+2\pi k/q$ and $\theta_0$ is a reference
angle. It is implicit in the definition (\ref{QFT}) that the
Hilbert space is of finite dimension $q$. The states
$|\theta_k\rangle$ form an orthonormal set and in addition the
projector over the subspace of phase states is
$\sum_{k=0}^{q-1}|\theta_k\rangle\langle \theta_k|=1_q$, where
$1_q$ is the unity operator. Given a state $|F\rangle$ one can
write a probability distribution $|\langle \theta_k|F\rangle|^2$
which may be used to compute various moments, e.g. expectation
values, variances. The key element of the formalism is that first
the calculations are done in the subspace of dimension $q$, then
the limit $q \rightarrow \infty$ is taken.\cite{Pegg89}.

We are now in position to define a quantum phase-locking
operator.\cite{PlanatJOB04} Our viewpoint has much to share with
the classical phase-locking problem as soon as one reinterpret the
fraction $\frac{k}{q}$ in (\ref{QFT}) as arising from the resonant
interaction between two oscillators and the denominator $q$ as a
number which defines the resolution of the experiment. From now we
emphasize such phase states $|\theta'_k\rangle$ which satisfy
phase-locking properties and we impose the coprimality condition
\begin{equation}
(k,q)=1. \label{coprime}
\end{equation}
The quantum phase-locking operator is defined as
\begin{equation}
\Theta_q^{\rm{lock}}=\sum_{k}'\theta_k |\theta'_k\rangle\langle
\theta'_k|,\label{Qlock}
\end{equation}
with $\theta_k=2\pi\frac{k}{q}$
and the notation $\sum '$ means summation from $0$ to $q-1$ with
$(k,q)=1$. Using (\ref{QFT}) and (\ref{coprime}) in (\ref{Qlock})
one obtains
\begin{equation}
\Theta_q^{\rm{lock}}= \frac{1}{q}\sum_{n,l}c_q(n-l)|n\rangle\langle
l|, \label{projector}
\end{equation}
where the range of values of $n,l$ is from $0$ to $\phi(q)$, and
$\phi(q)$ is the Euler totient function. The coefficients in front
of the outer products $|n\rangle\langle l|$ are the so-called
Ramanujan sums
\begin{equation}
c_q(n)=\sum_{k}' \exp(2i\pi \frac{k}{q}
n)=\frac{\mu(q_1)\phi(q)}{\phi(q_1)},~~\rm{with}~q_1=q/(q,n).
\label{Rsums}
\end{equation}
In the above equation $\mu(q)$ is the M\"obius function, which is
$0$ if the prime number decomposition of $q$ contains a square,
$1$ if $q=1$ and $(-1)^K$ if $q$ is the product of $K$ distinct
primes. Ramanujan sums are relative integers which are
quasi-periodic versus $n$ with quasi-period $\phi(q)$, and
aperiodic versus $q$ with a type of variability imposed by the
M\"obius function. Ramanujan sums have been used for signal
processing of low frequency noise.\cite{PRE02} In the Ramanujan
sum expansion there is a modified Mangoldt function $b(n)$ which
is the dual of M\"obius function
\begin{equation}
b(n)=\frac{\phi(n)}{n}\Lambda(n)=\sum_{q\ge
1}\frac{\mu(q)}{\phi(q)}c_q(n). \label{dual}
\end{equation}
This illustrates that many \lq\lq interesting" arithmetical
functions carry the structure of prime numbers. We mention the
relation $d \ln\zeta(s)/ds=\sum_{n\ge 1}\frac{\Lambda(n)}{n^s}$,
but there is also the relation $1/\zeta(s)=\sum_{n\ge
1}\frac{\mu(n)}{n^s}$. There is a well known
formulation\cite{FNL01}  of Riemann hypothesis from the summatory
M\"obius function $\sum_{n=1}^t \mu(n)=O(t^{1/2+\epsilon})$,
whatever $\epsilon$.

Given a state $|\beta\rangle$ one can calculate the expectation
value of the quantum phase-locking operator as
\begin{equation}
\langle \Theta_q^{\rm{lock}}\rangle=\sum_k' \theta_k|\langle
\theta'_k|\beta \rangle|^2. \label{expec}
\end{equation}
If one uses the finite form of Susskind-Glogower phase states
(\ref{Susskind}) and a real parameter $\beta$
\begin{equation}
|\beta\rangle =\frac{1}{\sqrt{q}}\sum_{n=0}^{q-1}\exp(i n
\beta)|n\rangle, \label{CohState}
\end{equation}
\begin{figure}[htbp]
\centering{\resizebox{7cm}{!} {\includegraphics{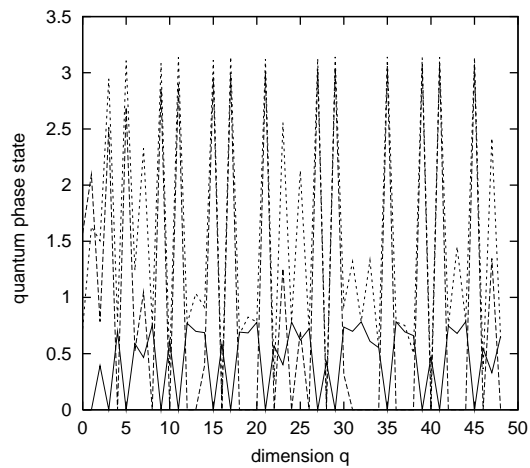}}}
\caption{Oscillations in the expectation value (\ref{expec2}) of
the locked phase at $\beta=1$ (dotted line) and their squeezing at
$\beta=0$ (plain line). The brokenhearted line which touches the
horizontal axis is $\pi \Lambda(q)/\ln q$} \label{Fig2}
\end{figure}
the expectation value of the locked phase becomes
\begin{equation}
\langle
\Theta_q^{\rm{lock}}\rangle=\frac{\pi}{q^2}\sum_{n,l}c_q(l-n)\exp(i\beta(n-l)).
\label{expec2}
\end{equation}
For $\beta=1$ it is found that $\langle
\Theta_q^{\rm{lock}}\rangle$ has the more pronounced peaks are at
such values of $q$ which are powers of a prime number. It can be
approximated by the normalized Mangoldt function $\pi
\Lambda(q)/\ln q$ as shown on Fig. 2. For $\beta=0$ the
expectation value of $\langle \Theta_q^{\rm{lock}}\rangle$ is much
lower. The parameter $\beta$ can be used to minimize the phase
uncertainty well below the classical value.\cite{PLA03}

Quantum phase-locking effect and its relation to prime number
theory has also been studied implicitely by Bost and
Connes.\cite{Bost95} Instead of an ad-hoc quantum phase operator
as (\ref{Pegg}) or (\ref{projector}), it is based on the
formulation of a dynamical system and its associated quantum
statistics. The dynamical system is first defined by an
Hamiltonian operator $H_0$ with eigenvalues equal to the
logarithms of integers
\begin{equation}
H_0|n\rangle= \ln n |n\rangle.
\end{equation}
Using the relations $\exp(-\beta_0 H_0)|n\rangle=\exp(-\beta_0\ln
n)|n\rangle=n^{-\beta_0}|n\rangle$, it follows that the partition
function of the model at the inverse temperature $\beta_0$ is
\begin{equation}
\mbox{Trace}(\exp(-\beta_0 H_0))=\sum_{n=1}^{\infty}
 n^{-\beta_0}=\zeta(\beta_0),
\end{equation}
where $\zeta(\beta_0)$ is the Riemann zeta function

In quantum statistical mechanics, given an observable Hermitian
operator $M$ one has the Hamiltonian evolution $\sigma_t(M)$
versus time $t$
\begin{equation}
\sigma_t(M)=e^{itH_0}Me^{-itH_0},
\end{equation}
and the expectation value of $M$ is the Gibbs state \\
$\mbox{Gibbs}(M)=\mbox{Trace}(M \exp(-\beta_0
H_0))/\mbox{Trace}(\exp(-\beta_0 H_0)$.
\begin{figure}[htbp]
\centering{\resizebox{8cm}{!} {\includegraphics{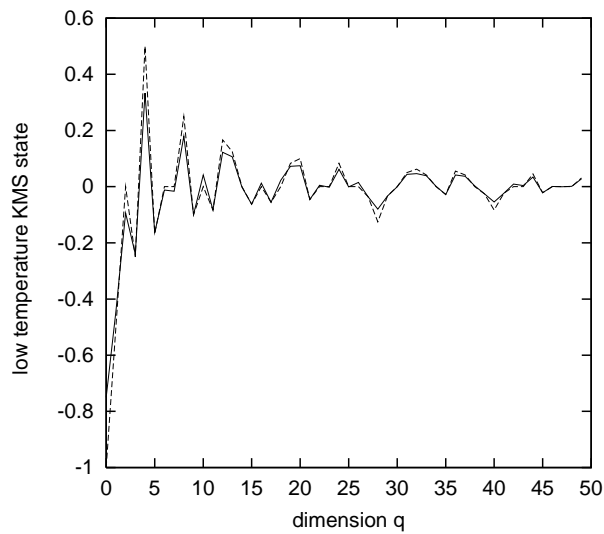}}}
\caption{Phase expectation value (\ref{KMS}) in Bost and Connes
model at the inverse temperature $\beta=3$ (plain line) in
comparison to the function $\mu(q)/\phi(q)$ (dotted line)}
\end{figure}
\begin{figure}[htbp]
\centering{\resizebox{8cm}{!} {\includegraphics{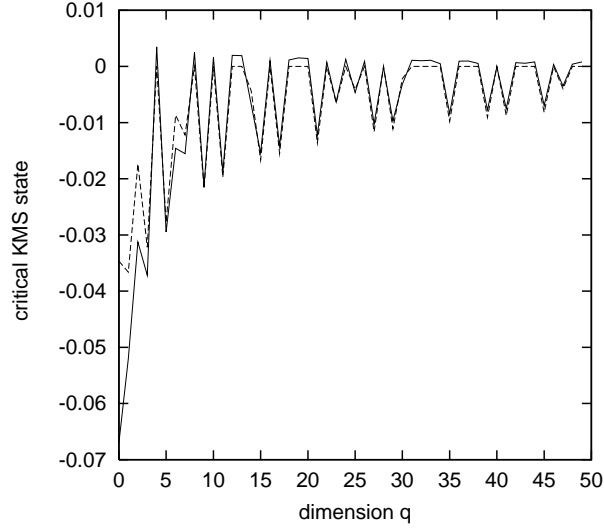}}}
\caption{Phase expectation value (\ref{KMS}) in Bost and Connes
model at the inverse temperature $\beta=1+\epsilon$, $\epsilon=0.1$
(plain line) in comparison to the function $-\Lambda(q)\epsilon/q$
(dotted line).}
\end{figure}
In Bost and Connes approach the observables belong to an algebra
of operators $\mu_a$ and $e_k$ which are defined by their action
on the occupation numbers $|n\rangle$ as
\begin{eqnarray}
&\mu_a|n\rangle =|an~ (\rm{mod}~ q)\rangle, \label{shift}\\
&e_k|n\rangle=\exp(\frac{2i \pi k n}{q})|n\rangle. \label{Qphase}
\end{eqnarray}
The first operator $\mu_a$ acts as a shift in the space of number
states; the second one $e_k$ is such that its action encodes the
individuals in the quantum Fourier transform (\ref{QFT}). Like in
the quantum phase-locking operator one uses the coprimality
condition (\ref{coprime}) to distinguish in (\ref{Qphase}) the
primitive roots of unity $\exp(2i \pi k/q)$, $(k,q)=1$. One can
show that there is a hidden symmetry group which is used to label
the elements of the algebra\footnote{This is the Galois group
$W=Gal(\mathcal{Q}^{\rm{cycl}}/\mathcal{Q})$ of the cyclotomic
extension on the field of rational numbers $\mathcal{Q}.$}. Using
the action of the group, the Gibbs state is replaced by the
so-called Kubo-Martin-Schwinger (or KMS) state. The system
exhibits a phase transition with spontaneous symmetry breaking at
the inverse temperature $\beta_0=1$ which corresponds to the
unique pole of the Riemann zeta function $\zeta(\beta_0)$. At low
temperature $\beta_0 >1$ one gets, after tricky calculations, the
expectation value of the phase operator which replaces
(\ref{expec2}) in the following form\cite{Bost95}
\begin{equation}
\mbox{KMS}(e_q^{(p)})=q^{-\beta_0}\prod_{\stackrel{p~
\rm{divides}~ q }{p~ \rm{prime}}}\frac{1-p^{\beta_0
-1}}{1-p^{-1}}.  \label{KMS}
\end{equation}
The KMS state is represented for two limiting cases, the low
temperature limit $\beta\gg 1$ (Fig. 3) and the critical case
$\beta=1+\epsilon$ (Fig. 4), with $\epsilon\simeq 0$. In these
limits one has respectively $\mbox{KMS}_{\beta\gg
1}(q)=\frac{\mu(q)}{\phi(q)}$ and $\mbox{KMS}_{1+\epsilon}\simeq
\frac{-\Lambda(q)\epsilon}{q}$. In the low temperature limit the
spectrum (\ref{dual}) corresponding to the Ramanujan sum expansion
of the modified Mangoldt function
$b(n)=\frac{\Lambda(n)\phi(n)}{n}=\Lambda(n)$ is recovered (see
(\ref{dual}). Close to the critical point $\beta=1+\epsilon$ the
oscillations are proportional to $\Lambda(q)\simeq b(q)$ and are
of very small amplitude due to the squeezing coefficient
$\epsilon$. A comparable squeezing effect was already observed in
the expectation value (\ref{expec}) of the quantum phase operator
(see Fig. 2).

Thus after the phenomenological model (\ref{flucgain}), the Bost
and Connes cyclotomic model also points to the Mangoldt function
as a source of low frequency fluctuations. In the last case the
model is associated to the spontaneous symmetry breaking and the
squeezing of phase oscillations at the critical KMS state
\footnote{The theory was also used as a model of time
perception.\cite{PlanatNQ04}}.

\section{Galois Phase-locking and Quantum Complementarity}

{\it But what is light really? Is it a wave or a shower of
photons? There seems no likelihood for forming a consistent
description of the phenomena of light by a choice of only one of
the two languages. It seems as though we must use sometimes the
one theory and sometimes the other, while at times we may use
either. We are faced with a new kind of difficulty. We have two
contradictory pictures of reality; separately neither of them
fully explains the phenomena of light, but together they do."}

(Albert Einstein and Leopold Infeld, The Evolution of Physics, p.
262).

More hints into quantum phase-locking may be discovered by deriving
a mathematical view of the complementarity principle. At the
conceptual level, two observables are complementary if precise
knowledge of one of them implies that all possible outcomes of
measuring the other one are equally probable. The eigenstates of
such complementary observables are non-orthogonal quantum states,
and in any attempt to distinguish between them, information gain is
only possible at the expense of introducing disturbance.

Mathematically speaking let $\mathcal{O}$ be an observable in a
Hilbert space of dimension $q$, $\mathcal{H}_q$, which is
represented by a Hermitian $q \times q$ matrix. Let us assume that
its real eigenvalues are multiplicity-free and its eigenvectors
$|b\rangle$ belong to an orthonormal basis $B$. Let $\mathcal{O}'$
be a (prepared) complementary observable with eigenvectors
$|b'\rangle$ in $B'$. If $\mathcal{O }$ is measured, then the
probability to find the system in the state $|b\rangle \in B$ is
given by $|\langle b|b'\rangle |^2= 1/q$. We here recall that two
orthonormal bases $B$ and $B'$ of $\mathcal{H}_q$ are mutually
unbiased precisely when $|\langle b| b'\rangle|^2=\frac{1}{q}$ for
all $b \in B$ and $b' \in B'$. It can be shown that in order to
fully recover the density matrix of a set of identical copies of a
quantum state, we need at least $q+1$ measurements performed on
complementary observables. As a matter of fact, the mathematical
implementation of the complementary principle lead us to the
search of completes sets of mutually unbiased bases (or MUBs for
short), a problem which has recently received a peculiar
attention.\cite{QuiPro}

In dimension $q=2$, eigenvectors of ordinary Pauli spin matrices
(i.e. in dimension $q=2$) provide the best known example of a
complete set of MUBs. It has been shown that in dimension $q=p^m$
which is the power of a prime $p$, the complete sets of mutually
MUBs result from Fourier analysis over a Galois field
$\mathbf{F}_q$ (in odd characteristic $p$)\cite{Wootters89} or of
a Galois ring $\mathbf{R}_{4^m}$( when $p=2$) (see
Refs.~\refcite{Klapp03} and \refcite{Planat05} for more details).
Complete sets of MUBs have an intrinsic geometrical
interpretation, and were for example related to discrete phase
spaces\cite{Wootters04bis} or finite projective
planes.\cite{Saniga,Saniga3}

The complete sets of MUBs in odd power dimension $q$ have a very
compact Fourier form\cite{Planat05}
\begin{equation}
|\theta_b^a\rangle=\frac{1}{\sqrt{q}}\sum_{n\in
\mathbf{F}_q}\psi(n)\kappa(an^2+bn)|n\rangle,~~a,b \in
\mathbf{F}_q. \label{MUB1}
\end{equation}
in which the coefficient in the computational base
$\{|0\rangle,|1\rangle,\cdots,|q-1\rangle\}$ represents the
product of an arbitrary multiplicative character $\psi(n)$ by an
arbitrary additive character $\kappa(yn)$, and where the
decomposition $y=a n+b$, has been used.

Let  fix a primitive root $g$ in $\mathbf{F}_q$, then every $n \in
\mathbf{F}_q$  is given by $n = g^s$ for some $s \in [0, q-2]$ and
then
\begin{equation}
\psi(n) = \omega_{q-1}^{k s},
\end{equation}
is a multiplicative character. Every multiplicative character can
be obtained in this way.

The additive characters are defined as
\begin{equation}
\kappa(x)=\omega_p^{tr(x)},~~\omega_p=\exp (\frac{2i\pi}{p}),~~x
\in \mathbf{F}_q.
\end{equation}
where $tr(x)=x+x^p+\cdots+x^{p^{m-1}}$ is the field theoretical
trace. It maps any element of $\mathbf{F}_q$ to an element in the
base field $\mathbf{F}_p$.

Eq.~(\ref{MUB1}) defines a set of $q$ bases (with index $a$) of
$q$ vectors (with index $b$). Using a property of Weil
sums\cite{Planat05} it is easily shown that, for $q$ odd, the
bases are orthogonal and mutually unbiased to each other and to
the computational base.

 The result of Wootters and Field corresponds to the trivial
multiplicative character $\psi_0=1$. Eq.~(\ref{MUB1}) also defines
phase states generalizing those written before in (\ref{QFT}). The
latter are recovered if one uses the trivial additive character
$\kappa=\kappa_0$ in (\ref{MUB1})\footnote{In (\ref{QFT}) we used
the (non prime) integer $p$ instead of $k$.}.

\subsection{The phase expectation value}

For the evaluation of the phase properties of a general pure state
of an electromagnetic field mode in the Galois number field we
proceed as in Sect. \ref{Quantum}. We consider the pure state of
the form (\ref{CohState}), and we sketch the computation of the
probability distribution $S=|<\theta_b|\beta>|^2$ and phase
expectation value $<\Theta_{\rm{Gal}}>=\sum_{b \in
\mathbf{F}_q}\theta_b|<\theta_b|\beta>|^2$. We recall that
$\theta_b=2\pi b/q $ (the upper index $a$ for the base is implicit
and we discard it for simplicity). The probability distribution
reads explicitly as\cite{Planat05}
\begin{equation}
\frac{1}{q^2} [\sum_{n \in \mathbf{F}_q} \psi(-n) \exp(i n
\beta)\kappa (-an^2-bn)][\sum_{m \in \mathbf{F}_q} \psi(m) \exp(-i
m \beta)\kappa (am^2+bm)]. \label{probability}
\end{equation}
Strictly speaking the phase state defined in (\ref{CohState}) is
defined only in prime dimension $q=p$. In non-prime dimensions the
product in the exponential $n \beta$ of (\ref{CohState}) is not in
$\mathbf{F}_p$.

For prime dimensions $p$ the phase probability distribution reads
\begin{eqnarray}
&S=\frac{1}{p^2}\sum_{n=1}^p \psi(n)\exp(2i\pi (\gamma n
+2an/p))\nonumber\\ &\times \sum_{n=1}^p \bar{\psi}(m))\exp(-2i\pi
(\gamma
m +2am/p))\nonumber\\
&=\frac{1}{p^2}\sum_{n,m=1}^p
\psi(n)\bar{\psi}(m)\exp(2i\pi(\gamma
(n-m)+a(n-m)(n+m)/p))\nonumber \\
&=\frac{1}{p^2}\sum_{n=1}^p\sum_{k=n-p}^{n-1}\psi(n)\bar{\psi}(n+k)
\exp(2i\pi \gamma k+a k (2n+k)/p)\nonumber\\
&=\frac{1}{p^2}\sum_{k=-p+1}^{p-1}\exp(2i\pi\gamma k)
T(k)\nonumber.
\end{eqnarray}
where we used the notation $\gamma=-\beta/2\pi+b/p$. In the last
but one equality above we put $k=n-m$ and in the last equality we
changed the order of summation, pulling out the $\gamma$-dependant
factor outside the $n$ summation. The inner sum equals
\begin{equation}
T(k)=\sum_{n=\rm{max}\{1,1-k\}}^{\rm{min}\{p-1,p-1-k\}}
\psi(n)\bar{\psi}(n+k) \exp(2i\pi ak(2n+k)/p). \label{incomplete}
\end{equation}

Now, if $k=0(\rm{mod}~ p)$ (which may happen only for $k=0$ in the
above range) the inner sum is trivial and is equal to $p$. For
other values of $k$ the inner sum is at most of absolute value
$O(p^{1/2}\ln p)$ by the Weil bound of incomplete
sums\cite{Lidl83,Shparlinski}(note that the factor involving
$\gamma$ is now gone from the sum over $n$).

Hence
\begin{equation}
|S(k)|=\frac{1}{p^2}O(1.p+p.p^{1/2}\ln p)=O(p^{-1/2}\ln p).
\label{bound_on_S}
\end{equation}
We get $|S|\simeq 0.63$ at p=3 and $0.49$ at $p=7$. Then it
decreases slowly with increasing dimension $p$.

The phase expectation value reads
\begin{equation}
<\Theta_{\rm{Gal}}>=\frac{2
\pi}{p^3}\sum_{k=-p+1}^{p+1}\exp(-i\beta k)T(k)\sum_{b=1}^p b
\exp(2i \pi k b/p).
\end{equation}
The partial sums in the above equation can be evaluated as
$p(p+1)/2$ for $k=0$ and otherwise
\begin{equation}
U=\sum_{b=1}^pb
\epsilon^b=\epsilon(1+2\epsilon+3\epsilon^2+\cdots+p\epsilon^{p-1})=\epsilon[\frac{1-\epsilon^p}{(1-\epsilon)^2}
-\frac{p \epsilon^p}{1-\epsilon}]=\frac{\epsilon p}{\epsilon-1},
\end{equation}
where we introduced $\epsilon=\exp(2i\pi k/p)$ and we made use of
the relation $\epsilon^p=1$. Easy calculations lead to
\begin{equation}
|U|=\frac{p}{2|\sin(2k \pi/p)|}.
\end{equation}
An estimate of the phase expectation value can be obtained as
follows.\cite{Iwaniec04} Let $r_k$ be the smallest (by absolute
value) residue of $2k (\rm{mod}~p)$. Then $|\sin( 2\pi k/p)| =
\sin(\pi |r_k|/p) \ge 2p|r_k|/\pi$ since $\sin (x) \ge 2x/\pi$ for
$0 \le x \le \pi/2$. We now define $r_k = 1$ for $k = 0$. Thus, we
now have $U(k) = O(p^2/r_k)$ for any $k$. Therefore $\sum_{k =
-p+1}^{p-1} |U(k)| = O(p^2 \sum_{k = -p -1}^{p+1} 1/r_k)$.

When $k$ runs between $-p-1$ and $p+1$, with $k \ne 0$, $r_k = 2 k
(\rm{mod}~ p)$ takes each value in the range $[0,(p-1)/2]$ no more
than $4$ times. The contribution from $k = 0$ is simply $1$. So
the contribution is $O(p^2 \sum_{k = -p -1}^{p+1} 1/r_k) = O(p^2
\sum_{s = 1}^{(p-1)/2} 1/s) = O(p^2 \log p)$.

As a result the phase expectation value is bounded by
\begin{eqnarray}
&|<\Theta_{\rm{Gal}}>|=\frac{2\pi}{p^3}O(1.p.
\frac{p(p+1)}{2}+p.p^{1/2}\ln p.p^2 \ln p)\\
\nonumber &= O(1+p^{1/2}(\ln p)^2), \label{expect}
\end{eqnarray}
which is a diverging quantity.

When $\psi(n)$ is a trivial character equal to $1$, the estimate
on incomplete Gauss sums (\ref{incomplete}) is replaced by the
better bound $O(p/\min {(k,p-k)})$ . It follows that
$S=O(p^{-1}\ln p)$\footnote{However one way to squeeze the bound
on the phase expectation value to $\pi$ is to project on the state
$| \beta \rangle=|0\rangle$ in (\ref{CohState}) as it was done in
Fig. 2 in the context of \lq\lq Ramanujan-type" phase states .}.

Here is the proof. When $\psi(n)$ is trivial we have $T(k) =
\sum_{n = N}^M \exp( 4 i\pi ak n/p)$ where $N$ and $M$ the lower
and upper limits in (\ref{incomplete}). Assume that $k \ne 0$.
Then $T(k)=(\exp( 4 i \pi ak N/p)-\exp( 4 i\pi ak M/p))/(1-exp( 4i
\pi ak /p))$. Estimating the numerator as $2$, we get $|T(k)| \le
2/|1-exp( 4i \pi ak /p)| = 1/|\sin( 2\pi ak/p)|$.

Let now use the notation $s_k$ for the smallest (by absolute
value) residue of $2ak (\rm{mod}~p)$. Therefore $\sum_{k =
-p+1}^{p-1} |T(k)| = O(p \sum_{k = -p -1}^{p+1} 1/s_k)=O(p \ln p)$
and $S(k)=O(p^{-1} \ln p)$ as expected.

\subsection{Phase variance}

The phase variance can be written as
\begin{equation}
<\Delta\Theta_{\rm{Gal}}^2>=\sum_{b \in \mathbf{F}_q}(\theta_b-
<\Theta_{\rm{Gal}}>)^2|<\theta_b|f>|^2.
 \label{variance}
\end{equation}
For the first term one gets
\begin{equation}
\sum_{b=1}^p \theta_b^2|<\theta_b|\beta>|^2=\frac{4
\pi^2}{p^4}\sum_{k=-p+1}^{p+1}\exp(-i\beta k)T(k)\sum_{b=1}^p b^2
\exp(2i \pi k b/p).
\end{equation}
The partial sums in the above equation can be evaluated as
$\frac{p^3}{3}$ for $k=0$ and otherwise
\begin{equation}
V=\sum_{b=1}^p b^2 \epsilon^b=\epsilon \frac{d}{d
\epsilon}(\frac{p}{\epsilon -1})
=\frac{-p\epsilon}{(\epsilon-1)^2}=\frac{p^3}{4\sin^2(\pi k/p)}.
\end{equation}
Using the same type of reasoning than in the last section
\begin{eqnarray}
&\frac{4\pi^2}{p^4}O(1.p. \frac{p^3}{3}+p.p^{1/2}\ln p.p^3\ln p
=O(1+3 p^{1/2}(\ln p)^2).
\end{eqnarray}

The second term is
\begin{eqnarray}
<\Theta_{\rm{Gal}}>^2 \sum_{b=1}^p |<\theta_b|\beta>|^2= \frac{1
}{p^2}\sum_{k=-p+1}^{p+1}\exp(-i\beta k)T(k)\sum_{b=1}^p \exp(2i
\pi k b/p).
\end{eqnarray}
The inner sum equals $p$ if $k=0$ and $0$ otherwise, so that the
whole contribution os $O(1)$.

Finally the third term is
\begin{equation}
-2<\Theta_{\rm{Gal}}>\sum_{b=1}^p \theta_b
|<\theta_b|\beta>|^2=-2<\Theta_{\rm{Gal}}>^2.
\end{equation}
Using (\ref{expect}) the absolute value is is bounded by
\begin{equation}
O(1+p (\ln p)^4).
\end{equation}
All estimates of the contributing terms in the variance are
diverging with $p$.


\section{Maximally entangled states}

By definition entangled states in $\mathcal{H}_q$ cannot be
factored into tensorial products of states in Hilbert spaces of
lower dimensions. There is an intrinsic relation between MUBs and
maximal entanglement.

Generalized Bell states may be defined using the multiplicative
Fourier transform (\ref{Pegg}) applied to the tensorial products
of two qudits,
\begin{equation}
|\mathcal{B}_{u,k}\rangle=\frac{1}{\sqrt{q}}\sum_{n=0}^{q-1}\omega_q^{k
n}|n,n+u\rangle. \label{FourierEntang}
\end{equation}
Also these states are both orthonormal, $\langle
\mathcal{B}_{u,k}|\mathcal{B}_{u',k'} \rangle =\delta_{u
u'}\delta_{kk'}$, and maximally entangled,
$trace_2|\mathcal{B}_{u,k}\rangle \langle\mathcal{B}_{u,k}|
=\frac{1}{q}I_q$. A more general class of maximally entangled
states is obtained using the Fourier expansion (\ref{MUB1}) over
$\mathbf{F}_q$ ($q$ odd) as follows
\begin{equation}
|\mathcal{B}_{u,b}^a\rangle=\frac{1}{\sqrt{q}}\sum_{n=0}^{q-1}\omega_p^{tr[(a
 n + b)n ]}|n,n+ u\rangle ~.
\label{entangledGalois}
\end{equation}
In general, for $q$ a power of a prime, starting from
(\ref{entangledGalois}) one obtains $q^{2}$ bases of $q$ maximally
entangled states. Each set of the $q$ bases (with $u$ fixed) has
the property of mutual unbiasedness. Similarly, for sets of
maximally entangled $m$-qubits one uses the Fourier transform over
Galois rings.\cite{Planat05}

\section{Conclusion}

The phase relation between a single piece and the whole system is
strongly contextual, but the working mathematics is amazingly
universal. In an electronic phase-locked loop  we found that the
position of mode-locked zones is controlled by the arithmetic of
irreducible fractions (\ref{confrac})-(\ref{beat}), and the
strength of lockings is related to prime number theory via the
Mangoldt function (\ref{flucgain}). We also developed two
different approaches of phase-locking within the context of
quantum optics. One of them uses a discrete phase operator
(\ref{Qlock}), and the phase expectation value also relates to
prime number theory via Ramanujan sums (see (\ref{expec})). In a
more sophisticated form it is linked to the Riemann zeta function
(see (\ref{KMS})). In a second approach, the phase locked states
are properly defined Fourier transforms over a Galois field (see
(\ref{MUB1})). They connect to mutually unbiased bases, which
appears in the mathematical formulation of quantum
complementarity. Incomplete Gauss sums are at the kernel of phase
variability in this case.

\section*{Acknowledgements}
The author acknowledges Igor Shparlinski for its tremendous
contribution in the last section of the manuscript.

\end{document}